\pdfoutput=1

\documentclass[11pt,twoside,a4paper,cmspaper,final,collab]{cms-tdr}
\def\svnVersion{493157P}\def\cmsCernNoTag{CERN-EP-2018-202}\def\cmsCernDate{\today}\def\cmsMessage{Published in Physics Letters B as \href{http://dx.doi.org/10.1016/j.physletb.2019.01.069}{\doi{10.1016/j.physletb.2019.01.069.}}}

\begin{document}\cmsNoteHeader{EXO-17-008}

\hyphenation{had-ron-i-za-tion}
\hyphenation{cal-or-i-me-ter}
\hyphenation{de-vices}
\RCS$HeadURL: svn+ssh://svn.cern.ch/reps/tdr2/papers/EXO-17-008/trunk/EXO-17-008.tex $
\RCS$Id: EXO-17-008.tex 493147 2019-03-31 15:00:20Z swmukher $

\newlength\cmsTabSkip\setlength{\cmsTabSkip}{1ex}

\cmsNoteHeader{EXO-17-008}

\title{Search for a \PWpr boson decaying to a $\Pgt$ lepton and a neutrino in proton-proton collisions at $\sqrt{s} = 13\TeV$}

\date{\today}

\abstract{
A search for a new high-mass resonance decaying to a $\Pgt$ lepton and a neutrino
is reported.
The analysis uses proton-proton collision data collected by the CMS experiment at the LHC at $\sqrt{s}=13\TeV$,
corresponding to an integrated luminosity of 35.9\fbinv.
The search utilizes hadronically decaying $\Pgt$ leptons.
No excess in the event yield is observed
at high transverse masses of the $\Pgt$ and missing transverse
momentum.
An interpretation of results within the sequential standard model excludes \PWpr boson masses below 4.0\TeV
at 95\% confidence level. Existing limits are also improved
on models in which the \PWpr boson decays preferentially to fermions of the third generation.
Heavy \PWpr bosons with masses less than 1.7--3.9\TeV, depending on the
coupling in the non-universal G(221) model, are excluded at 95\% confidence level. These are the most stringent limits on this model
to date.
}
\hypersetup{%
pdfauthor={CMS Collaboration},%
pdftitle={Search for a \PWpr boson decaying to a tau lepton and a neutrino in proton-proton collisions at sqrts =13 TeV},%
pdfsubject={CMS},%
pdfkeywords={CMS, physics, tau, MET}}

\maketitle

\section{Introduction}

New charged heavy gauge bosons, generally referred to as \PWpr bosons,
are predicted by various extensions of the standard model (SM).
An example is the sequential standard model (SSM)~\cite{Altarelli:1989ff},
featuring an extended gauge sector, which is often used as a benchmark model.
Lepton universality holds in the SSM; however,
there exist models without this assumption.
Nonuniversal gauge interaction models (NUGIMs)~\cite{Chiang:2009kb,Li:1981nk,Muller:1996dj,Malkawi:1996fs,Hsieh:2010zr}
predict an enhanced \PWpr boson branching fraction to the third generation fermions.
In this approach, the high top quark mass is associated with the large
vacuum expectation value of the corresponding Higgs field.

\begin{figure}[htp]
\centering
\includegraphics[width=0.49\textwidth]{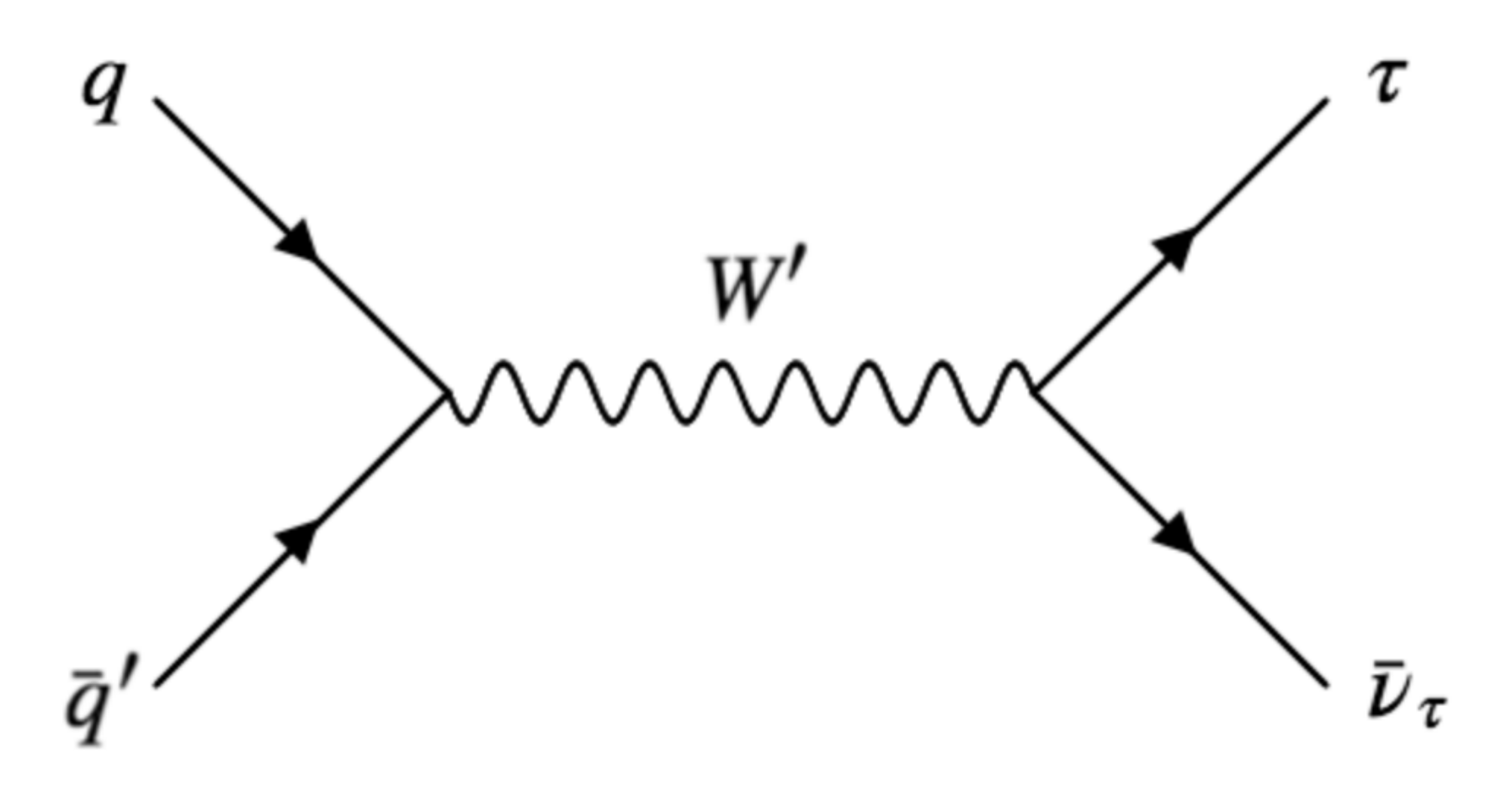}
\caption{Leading order Feynman diagram of the expected signal process $\PWpr \to \Pgt \Pgn$.}
\label{fig:feynman}
\end{figure}

The analysis presented in this Letter searches for $\PWpr \to \Pgt \Pgn$ events, where the $\Pgt$ lepton decays hadronically.
The leading order Feynman diagram is shown in Fig.~\ref{fig:feynman}. In this Letter, the symbol \tauh will be used
to denote the visible part of the hadronic decay of the $\Pgt$, which is reconstructed as a $\Pgt$ jet in the detector.
The hadronic decays of the $\Pgt$ lepton are
experimentally distinctive because they result in low charged-hadron multiplicity,
unlike jets originating from the hadronization of partons produced in the hard scattering process,
which have
high charged-hadron multiplicity.
The signature of a \PWpr boson event is similar
to that of a \PW\ boson event in which the \PW\ boson is produced ``off-shell'' with a high mass.

Searches for a \PWpr boson decaying to a $\Pgt$ lepton and a neutrino have been
performed previously by the CMS~\cite{Khachatryan:2015pua}
and ATLAS~\cite{Aaboud:2018vgh} collaborations at the CERN LHC.
Searches for a \PWpr boson have been performed also in $\Pe+\ptmiss$, $\Pgm+\ptmiss$~\cite{Sirunyan:2018mpc,Aaboud:2017efa},
\PW\PZ~\cite{Sirunyan:2018ivv,Aaboud:2017eta},
\cPq\cPq$^{\prime}$~\cite{Sirunyan:2016iap,Aaboud:2017yvp} and \cPqt\cPqb~\cite{Sirunyan:2017vkm,Aaboud:2018juj} channels.
The ATLAS experiment has excluded an SSM \PWpr for masses below 3.7\TeV in the $\tauh+\ptmiss$ channel.
The CMS experiment has excluded an SSM \PWpr for masses below 5.2\TeV in the combination of electron and muon
channels. This Letter describes
a search for a \PWpr boson in the $\tauh+\ptmiss$ channel using proton-proton~(\Pp\Pp) collisions
collected in 2016 at a center-of-mass energy of 13\TeV. The data set corresponds to an integrated luminosity of 35.9\fbinv.
The results are interpreted in the context of two models,
the SSM and the NUGIM.

\section{Physics models}

\subsection{The sequential standard model \PWpr boson}
\label{sec:models-ssm}

In the SSM, the \PWpr boson is a heavy analog of the \PW\ boson. It is a resonance with
fermionic decay modes and branching fractions similar to those of the SM \PW\ boson, with the
addition of the decay $\PWpr\to \cPqt\cPqb$, which becomes relevant for \PWpr boson masses larger than
180\GeV. If the \PWpr boson is heavy enough to decay to top and bottom quarks, the SSM
branching fraction for the decay $\PWpr \to \Pgt \Pgn$ is 8.5\%~\cite{Altarelli:1989ff}. Under these
assumptions, the relative width $\Gamma/M$ of the \PWpr boson is $\sim$3.3\%.
With increasing mass, a growing fraction of events are produced off-shell and shifted to lower mass values.
Assuming events within a window of $\pm$10\% around the actual mass to be on-shell, the off-shell fractions are approximately
9, 22 and 66\% for $\PWpr$ masses of 1, 3 and 5\TeV, respectively. Decays into
\PW\PZ depend on the specific model assumptions and are usually considered to be suppressed in the SSM, as assumed by
the current search.

In accordance with previous analyses, it is assumed that there is no interference between the production of the new
particle and the production of the SM \PW\ boson.
Such an absence of interference would occur,
for example, if the \PWpr
interacts via V+A coupling~\cite{Boos:2006xe}.

Signal events for the SSM \PWpr boson are simulated at leading order (LO) with \PYTHIA 8.212~\cite{Sjostrand:2014zea}, using
the NNPDF~2.3~\cite{Ball:2011mu,Ball:2012cx} parton distribution function (PDF) set
and tune CUETP8M1~\cite{Skands:2014pea}.
The \PWpr samples are normalized to next-to-next-to-leading-order (NNLO) cross sections from \FEWZ~\cite{Gavin:2012sy,Li:2012wna}.

\subsection{Coupling strength}
The \PWpr boson coupling strength, $g_{\PWpr}$, is given in terms of the SM weak coupling
strength $ g_{\PW} = e/\swsq \approx 0.65 $.
Here, $\theta_{\PW}$ is the weak mixing angle.
If the \PWpr is a heavier copy of the SM \PW\ boson, their coupling ratio is $g_{\PWpr}/g_{\PW}=1$ and the
SSM \PWpr theoretical cross sections, signal shapes, and widths apply.
However, different couplings are possible. Because of the dependence of the width of a particle on its coupling, and
the consequent effect on the transverse mass distribution, a limit can also be set on the coupling strength.
For this study, a reweighting procedure is used.
Some selected signal samples are simulated at LO with \MADGRAPH (version 1.5.11)~\cite{Alwall:2014hca}, for a range of coupling ratios $g_{\PWpr}/g_{\PW}$
from 0.01 to 3.
These signals exhibit different widths as well as
different cross sections. The generated distributions of the SSM \PYTHIA samples with $g_{\PWpr}/g_{\PW} = 1$ are reweighted to
take into account the decay width dependence, thus providing the appropriate reconstructed
transverse mass distributions for $g_{\PWpr}/g_{\PW} \ne 1$ . For $g_{\PWpr}/g_{\PW}=1$, the theoretical LO cross sections apply and this coupling
strength is used to compare the standard SSM samples with the reweighted ones, allowing
the reweighting method to be verified.

\subsection{Nonuniversal gauge interaction model}

Models with nonuniversal couplings predict an enhanced branching fraction for the third generation of fermions and
explain the large mass of the top quark. The nonuniversal gauge interaction models (NUGIMs) exhibit a $SU(2)_{\text{l}} \times SU(2)_{\text{h}} \times U(1)$
symmetry, and thus are often called G(221) models. Here the indices l and h refer to light and heavy, respectively.
The weak SM $SU(2)_{\text{W}}$ group is a low-energy limit of
two gauge groups, a light $SU(2)_{\text{l}}$ and a heavy $SU(2)_{\text{h}}$, which govern the couplings
to the light fermions of the first two generations and to the heavy fermions
of the third generation, respectively.
These two groups mix, resulting in an SM-like $SU(2)_{\text{W}}$ and an extended group $SU(2)_{\text{E}}$.
The $SU(2)_{\text{E}}$ extended gauge group gives rise
to additional gauge bosons such as a \PWpr. The mixing of the two gauge
groups
involves a
mixing angle of the extended group, $\theta_{\mathrm{E}}$, which modifies
the couplings to the heavy boson.
Consequently, the mixing modifies
\begin{figure}[htp]
\centering
\includegraphics[width=0.49\textwidth]{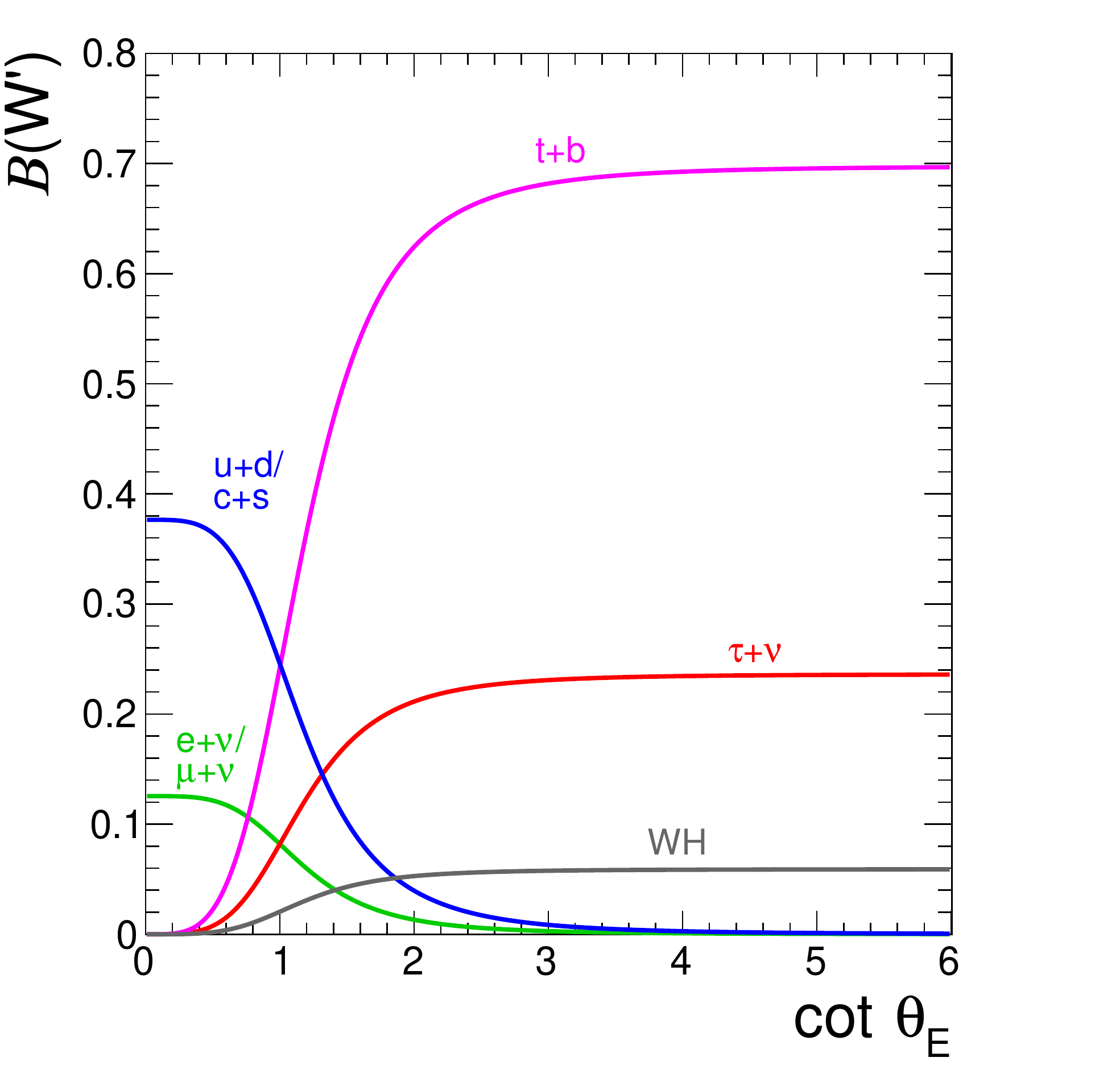}
\caption{
Branching fractions $\mathcal{B}(\PWpr)$ as a function of the mixing angle $\cot\theta_{\mathrm{E}}$, for \PWpr boson decays in the
NUGIM G(221) framework, as calculated in Refs.~\cite{Chiang:2009kb,Kim:2011qk,Edelhauser:2014yra}.
For $\cot\theta_{\mathrm{E}} = 1$ the values correspond to those in the SSM, rescaled to accommodate the \PW\PH decay channel.
}
\label{fig:branching}
\end{figure}
the production cross section
and, as illustrated in Fig.~\ref{fig:branching}, the branching fractions of the \PWpr.
For $\cot\theta_{\mathrm{E}} \gtrsim  3$ the \PWpr decays
predominantly to third generation fermions. The branching fraction to \PW\PH is smaller than the branching fraction to third generation fermions, as shown
in Fig.~\ref{fig:branching}.
For $\cot\theta_{\mathrm{E}}=1$ the branching fractions are the
same as those of the SSM, and the \PWpr boson couples
democratically to all fermions. For $\cot\theta_{\mathrm{E}} < 1$ the decays into light
fermions are dominant.

In the NUGIM G(221), the ratio of the couplings $g_{\PWpr}/g_{\PW}$ is related to the parameter $\cot\theta_{\mathrm{E}}$
by the following equation~\cite{Edelhauser:2014yra}:
\begin{equation}
    \Gamma_{\PWpr}=\Gamma_{\PWpr}^{\text{SSM}}
    \frac{(4+\frac{1}{4})\cot^2\theta_\mathrm{E}+8\tan^2\theta_\mathrm{E}}{12+\frac{1}{4}}
    =\Gamma_{\PWpr}^{\text{SSM}}\left(\frac{g_{\PWpr}}{g_{\PW}}\right)^2
\label{eq1}
\end{equation}
Because of this functional relationship, a reinterpretation of limits on coupling strength will yield limits on NUGIM G(221),
and thus it was not necessary to generate a signal sample for this model.

\section{The CMS detector}
The central feature of the CMS apparatus is a superconducting solenoid of 6\unit{m} internal
diameter, providing a magnetic field of 3.8\unit{T}. Within the solenoid volume are a silicon
pixel and strip tracker, a lead tungstate crystal electromagnetic
calorimeter (ECAL), and a brass and scintillator hadron calorimeter (HCAL), each composed of a barrel and two endcap sections.
Forward calorimeters extend the
pseudorapidity ($\eta$) coverage provided by the barrel and endcap detectors. Muons are detected in gas-ionization chambers embedded in the steel
flux-return yoke outside the solenoid.

The silicon tracker measures charged particles within the range $\abs{\eta} < 2.5$. It consists of 1440 silicon pixel and 15\,148 silicon strip
detector modules. For nonisolated particles with transverse momentum $1 < \pt < 10\GeV$ and $\abs{\eta} < 1.4$, the track resolutions are
typically 1.5\% in \pt and
25--90\mum in the transverse impact parameter and 45--150\mum in the longitudinal impact parameter.
The ECAL consists of 75\,848 lead tungstate crystals, which provide coverage of $\abs{\eta} < 1.48 $ in
a barrel region (EB) and $1.48 < \abs{\eta} < 3.0$ in two endcap regions (EE).
The HCAL is a sampling calorimeter, which utilizes alternating layers of brass as an absorber and
plastic scintillator as active material, covering the range $\abs{\eta} < 3$.
In the forward region, the calorimetric coverage is extended to $\abs{\eta}<5$ by a steel and quartz fiber Cherenkov hadron forward calorimeter.
Muons are measured in the range $\abs{\eta} < 2.4$, with detection planes made using three technologies: drift tubes, cathode strip chambers, and resistive
plate chambers. Events of interest are selected using a two-tiered trigger system~\cite{Khachatryan:2016bia}.

A more detailed description of the CMS detector, together with a definition of the coordinate system used and the relevant kinematic variables,
can be found in Ref.~\cite{Chatrchyan:2008zzk}.

\section{Background simulation}

The dominant SM background is the production of \PW+jets. This background is generated at LO using
\MGvATNLO version 2.2.2 with the MLM merging~\cite{Alwall:2014hca,Alwall:2007fs} and the NNPDF~3.0~\cite{Ball:2014uwa,Ball:2011mu,Ball:2012cx} PDF set
for on-shell \PW\ boson production and
using \PYTHIA~8.212 with the NNPDF~2.3 PDF set for off-shell production.
The differential cross section is reweighted as a function of the invariant mass of the SM \PW\ boson decay products,
incorporating NNLO quantum chromodynamics (QCD) and next-to-leading-order (NLO) electroweak (EW) corrections.
The effect with respect to the LO calculation corresponds to a correction factor ($K$ factor) for the \PW\ boson transverse mass
spectrum.
To combine the QCD and EW differential cross sections, two different mathematical approaches could be taken~\cite{Balossini:2009sa}:
an additive or a multiplicative combination. Their effects differ by around 5\%.
The $K$ factor assumed in this analysis is obtained by taking the additive combination as recommended by Ref.~\cite{Andersen:2014efa}
and the difference from the multiplicative combination is treated as a systematic uncertainty.
The $K$ factor is 1.15 at a \PWpr mass of 0.3\TeV and drops monotonically for higher masses down to 0.6 for a mass of 6\TeV.
The calculation uses the generators \FEWZ~3.1 and \textsc{mcsanc}~1.01~\cite{Bondarenko:2013nu} for the QCD and electroweak
corrections.

Other background processes are: $\PZ/\PGg* \to \ell \ell$ generated with \MGvATNLO version 2.3.2.2~\cite{Alwall:2014hca} with the NNPDF~3.0 PDF set,
diboson processes generated with \PYTHIA 8.212 and with the NNPDF~2.3 PDF set, and top quark
processes generated with \POWHEG~2.0~\cite{Frixione:2007vw,Nason:2004rx,Alioli:2010xd,Alioli:2009je,Re:2010bp,Frixione:2007nw} and the NNPDF~3.0 PDF set.
Background from jets that are falsely identified as $\tauh$ candidates is dominated by
$\PZ \to \Pgn\Pgn+\text{jets}$ events, which are simulated with \MGvATNLO version 2.3.2.2 and with the NNPDF~3.0 PDF set.

Parton fragmentation and hadronization
are performed with \PYTHIA~8.212 with the underlying event tune CUETP8M1.
The detector response is simulated using a detailed description of the CMS detector implemented with the \GEANTfour package~\cite{Agostinelli:2002hh}.
All simulated event samples are normalized to the integrated luminosity of the recorded data,
using the theoretical cross section values.
Additional pp collisions during the same bunch crossing (pileup)
is taken into account
by superimposing simulated minimum bias interactions onto all simulated events.
The simulated events are weighted so that the pileup distribution matches that of the data, with an average of about 27 interactions per bunch crossing.

\section{Reconstruction and identification of physics objects}

A particle-flow (PF) algorithm~\cite{CMS-PRF-14-001} is used to combine information from all CMS subdetectors in order to
reconstruct and identify individual particles in the event: muons, electrons, photons, and charged and neutral hadrons.
The resulting set of particles is used to reconstruct the $\tauh$ candidates, missing transverse momentum (\ptmiss), and jets.
The vector \ptvecmiss is defined as the negative vector \pt sum of all PF candidates reconstructed in the event. The magnitude of this
vector is referred to as \ptmiss. The raw \ptmiss value is modified to account for corrections to the
energy scale of all the reconstructed jets in the event~\cite{Khachatryan:2014gga}.
The jets are clustered using the anti-\kt jet finding algorithm~\cite{Cacciari:2008gp,Cacciari:2011ma}.
The reconstructed vertex with the largest value of summed physics-object $\pt^2$ is taken as the primary vertex.

Electrons~\cite{EGM-13-001,DPS-17-004} are reconstructed by matching energy deposits in the ECAL with track
segments in the inner tracker. Muon reconstruction~\cite{Sirunyan:2018fpa}
is performed by matching a track segment reconstructed in the inner tracker with
a track segment reconstructed in the muon detector and performing a global fit of the
charge deposits from the two track segments.

The $\tauh$ reconstruction in CMS starts from jets
clustered from PF candidates,
using the anti-\kt algorithm with a distance parameter of 0.4.
The $\tauh$ candidates are reconstructed using the hadrons-plus-strips algorithm~\cite{Khachatryan:2015dfa,TAU-16-002}, which is
designed to optimize the performance of $\tauh$ reconstruction and identification by considering
specific $\Pgt$ lepton decay modes. Individual $\tauh$ decay modes are reconstructed separately.
The signatures distinguished by the algorithm are: a single charged hadron, a charged hadron and up to two neutral pions,
and three charged hadrons.

Requiring $\tauh$ candidates to pass isolation requirements
reduces the $\text{jet} \to \tauh$ misidentification probability.
The multivariant-based (MVA-based) $\tauh$ identification discriminant combine isolation and other variables
with sensitivity to the $\Pgt$ lifetime, to provide the best possible discrimination for $\tauh$ decays against quark and gluon jets.
Hadronically decaying $\Pgt$ leptons in this analysis are required to satisfy the very loose working point of the MVA-based isolation~\cite{Sirunyan:2017khh}.
This working point has an efficiency of about 70\% for genuine $\tauh$, with about 0.4\% misidentification rate for quark- and gluon-initiated jets,
for a \pt range typical of $\tauh$ originating from a $\PWpr$ boson of mass of 2\TeV.
Isolated electrons have a high probability to be misidentified as $\tauh$ objects that decay to a
single charged hadron ($\text{h}^{\pm}$ or $\text{h}^{\pm}\Pgpz$).
Electrons can emit energetic bremsstrahlung photons as they traverse the material of the silicon tracker.
When this occurs, the electron and accompanying photons may be mistakenly reconstructed as a hadronically decaying $\Pgt$.
Muons can also be reconstructed as $\tauh$ objects in the $\text{h}^{\pm}$ decay mode.
The $\tauh$ candidates in this analysis are required to pass the loose working point of the antielectron discriminator, which
has an efficiency of about 85\% for genuine $\tauh$ events, and a misidentification rate of about 1.5\% for electrons.
The $\tauh$ candidates are further required to pass the loose working point of the antimuon discriminator, which
has an efficiency of $>99\%$ for genuine $\tauh$ events, with a misidentification rate of about 0.3\% for muons~\cite{Sirunyan:2018pgf,Sirunyan:2017khh}.

\section{Analysis strategy}
\label{sec:analysis}
The discriminating variable used in this analysis is the transverse mass, defined as follows:
\begin{equation}
\mT = \sqrt{2 \pt^{\Pgt}  \ptmiss [1-\cos\Delta\phi(\ptvec^{\Pgt},\ptvecmiss)]},
\end{equation}
where $\pt^{\Pgt}$ is the magnitude of the transverse momentum vector of the $\tauh$
candidate $\ptvec^{\Pgt}$, and $\Delta\phi$ is the difference in the azimuthal angle between $\ptvec^{\Pgt}$
and $\ptvecmiss$.

The strategy of this analysis is to select a heavy boson candidate decaying
almost at rest to a hadronic jet consistent with a $\tauh$ candidate and neutrinos, the latter manifesting themselves as \ptmiss.
Signal events are selected online with a $ \tauh+\ptmiss$ trigger that requires the \pt of the $\tauh$ candidate to be
greater than 50\GeV and the value of \ptmiss to be greater than 90\GeV. To ensure that the trigger is maximally
efficient for selected events, the offline selection requires one isolated $\tauh$ candidate to
have $\pt^{\Pgt}$ greater than 80\GeV and \ptmiss to be greater than 200\GeV.

Although there are two neutrinos in the final state, \ptmiss and the isolated $\tauh$ candidate are largely produced in
opposite directions, which helps to distinguish signal from background events especially those coming from QCD multijet production.
Two selection criteria exploit this behavior to reduce the background:
the ratio of the $\pt^{\Pgt}$ to \ptmiss is required to satisfy $0.7 < \pt^{\Pgt}/\ptmiss < 1.3$; and the
angle $\Delta\phi(\ptvec^{\Pgt}, \ptvecmiss)$ has to be greater than 2.4 radians.
Consequently, the lowest \mT value is about 300\GeV.
To avoid an overlap with the \PWpr boson search in the electron channel, events are rejected if they contain a loosely identified electron with
$\pt>20\GeV$ and $\abs{\eta}<2.5$, where the loose working point is $\approx$90\% efficient for real electrons.
For similar reasons, events containing a loosely identified muon with $\pt>20\GeV$ and $\abs{\eta}<2.4$ are not considered in this analysis, where the
loose working point is $>$99\% efficient for real muons.

\begin{figure*}[htb]
\centering
\includegraphics[width=0.49\textwidth]{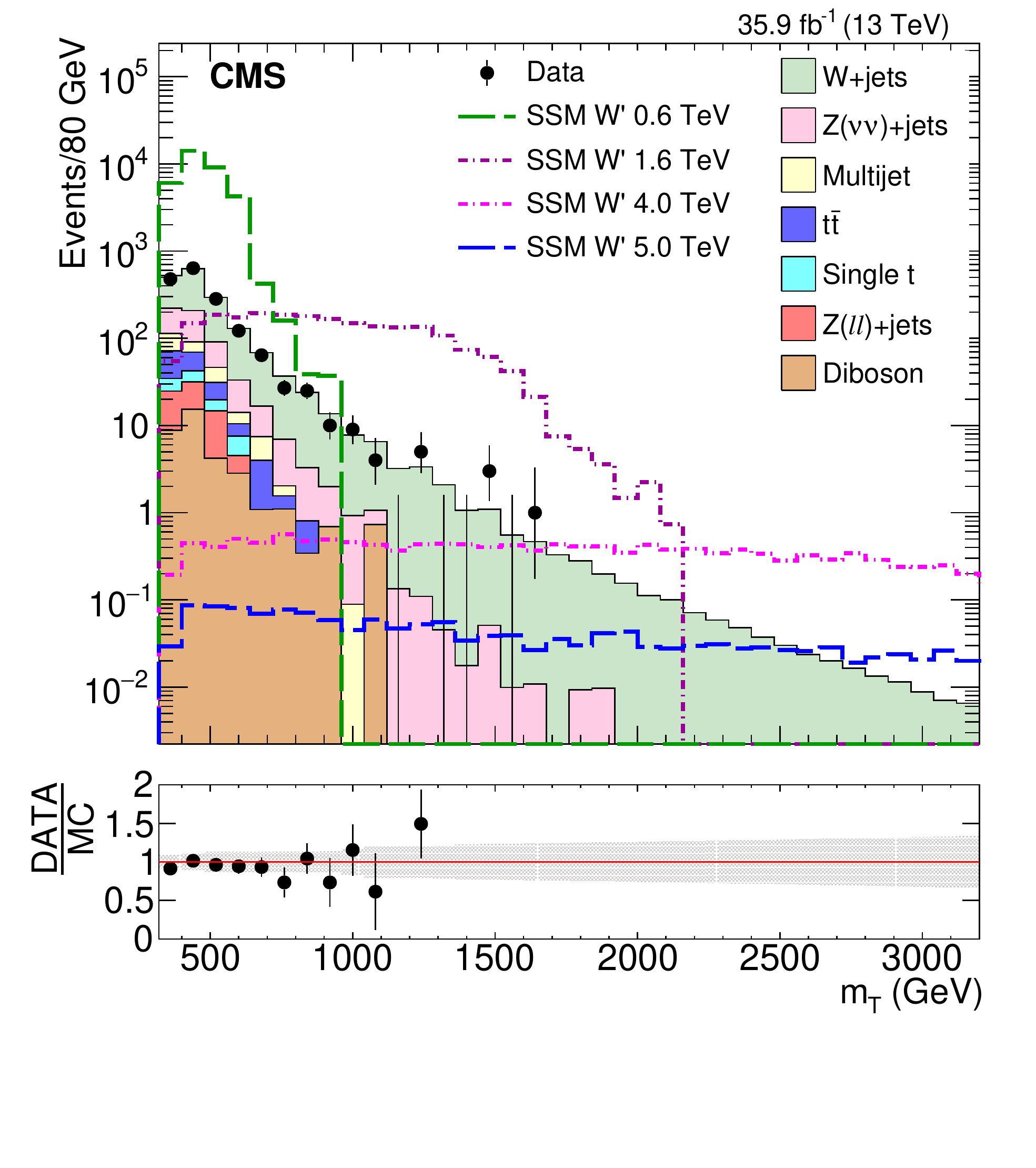}
\includegraphics[width=0.49\textwidth]{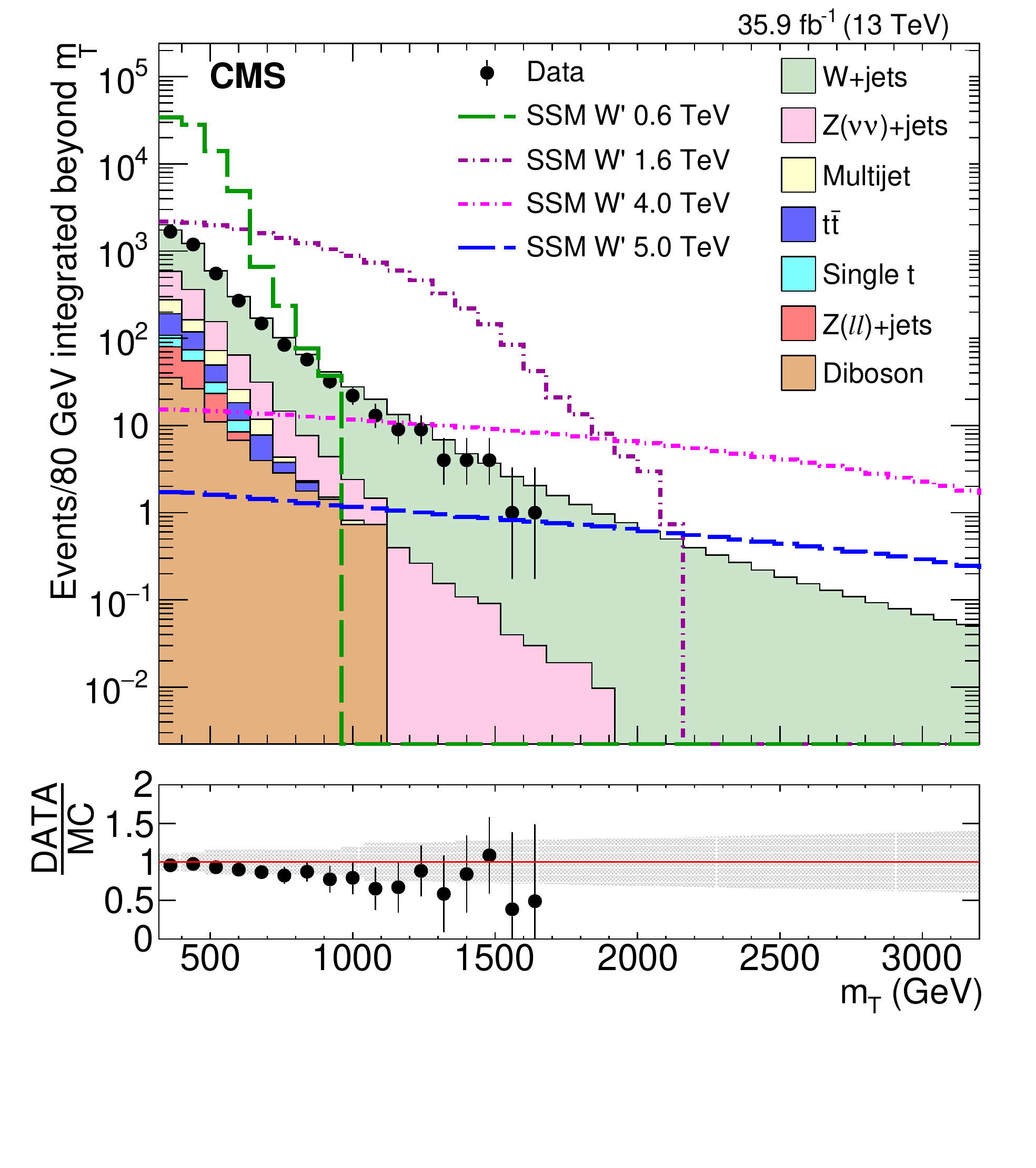}
\caption{
(Left) The \mT distribution after the final selection.
The black symbols with error bars show data, while the filled histograms represent the SM backgrounds.
Signal examples for SSM \PWpr bosons with masses of 0.6, 1.0, 4.0, and 5.0\TeV are shown with the open histograms.
(Right) The integral transverse mass distribution, where the value in each bin is equal to the number
of events with transverse mass equal to or greater than the left of the bin.
The lower panels show the ratio of data to prediction, and the gray band represents
the systematic uncertainties.
}
\label{fig:MT}
\end{figure*}

After all selections,
the \mT distributions for the observed data and expected background events
are presented in Fig.~\ref{fig:MT} (left).
Figure~\ref{fig:MT} (right) shows the integral distribution,
which is formed by filling each bin of the histogram with the sum of that bin and all following bins.
The systematic uncertainties, which are detailed in Section~\ref{sec:uncertainties}, are illustrated as a grey band in the lower panels of the plots.
The product of the signal efficiency and acceptance for SSM $\PWpr \to \Pgt \Pgn$ events depends on the \PWpr boson mass.
The total signal efficiency for the studied
range of $\mT>300\GeV$ varies from 14\% to about 24\% as $M_{\PWpr}$ increases from 1 to 3\TeV.
For higher \PWpr boson masses, events shift to lower \mT because of the increasing fraction of
off-shell production
(as shown in Fig.~\ref{fig:MT} for a few signal mass points).
For example, for a \PWpr boson with a mass of 5\TeV, the total signal efficiency is around 17\%.
The trigger threshold affects the signal efficiency in the low-mass range.
These efficiency values are obtained assuming the $\PWpr \to \Pgt \Pgn$ branching fraction to be unity.
The efficiency values are estimated using simulated events where the $\Pgt$ lepton decays hadronically.

\begin{figure}[htp]
\centering
\includegraphics[width=0.49\textwidth]{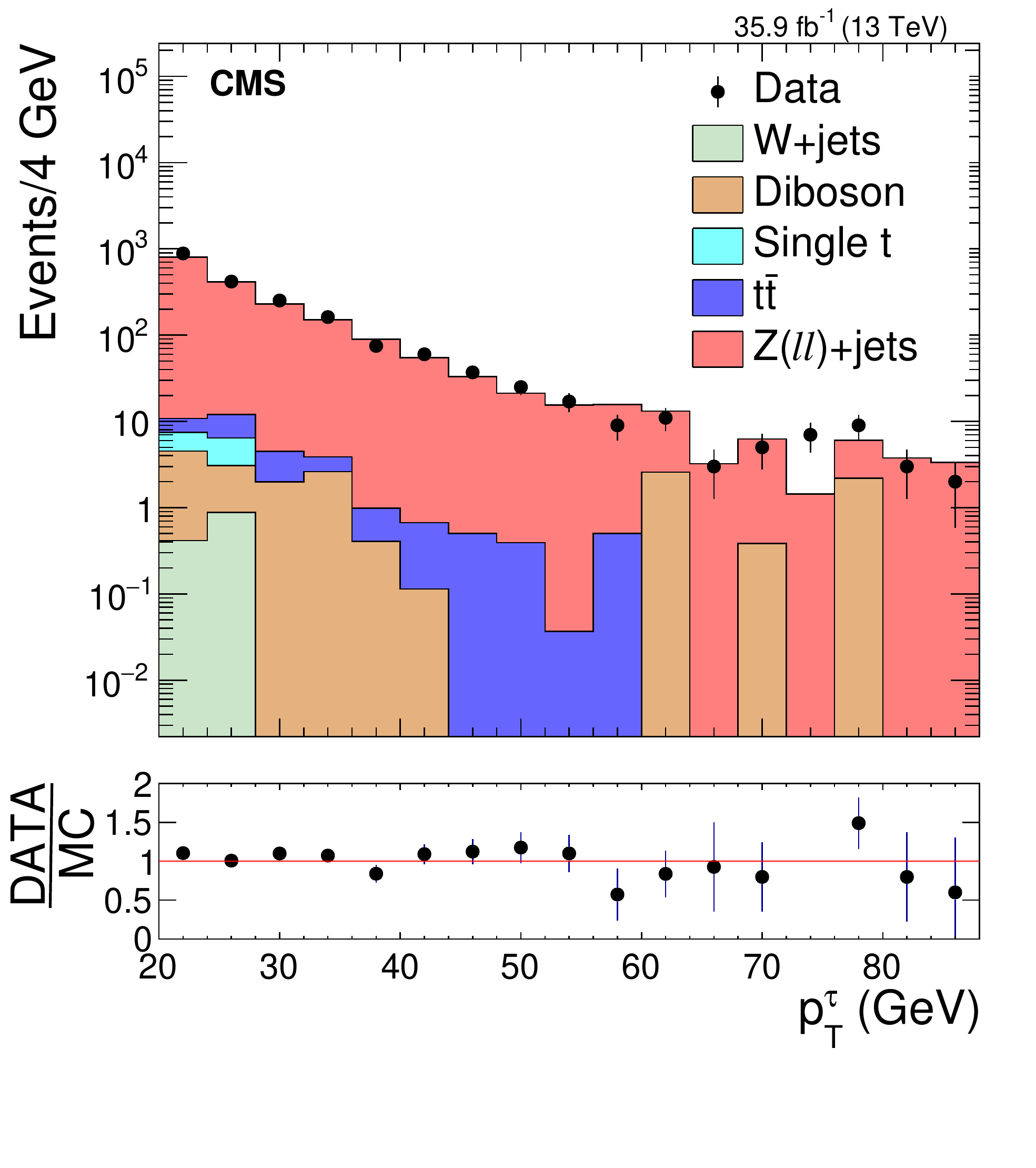}
\caption{Distribution of $\pt^{\Pgt}$ in the control region. The black symbols with error bars
show the data, while the histograms represent the SM backgrounds. The lower panel
shows the ratio of data to prediction.
}
\label{fig:CRplot}
\end{figure}

The dominant background is from the off-shell tail of the \mT distribution of
the SM \PW\ boson, and is obtained from simulation.
The background contributions from $\PZ(\to \Pgn\Pgn)+\text{jets}$ and QCD multijet events are also obtained from simulation.
These backgrounds primarily arise as a consequence of jets misidentified as $\tauh$ candidates. The contribution of QCD multijet background
is small compared
to $\PZ(\to \Pgn\Pgn)+\text{jets}$ in the signal region.
Following the strategy in Ref.~\cite{Sirunyan:2017hci}, to ensure that the misidentified $\Pgt$ background
is simulated properly, the agreement between data and simulation is checked in a control region dominated by
$\PZ(\to \Pgm\Pgm)+\text{jets}$ events, where a jet is misidentified as a $\tauh$ candidate.
The \ptmiss is recalculated excluding the muons from the Z decay in order to reproduce the \ptmiss distribution of $\PZ \to \Pgn \Pgn$ events.
Specifically, the control region is defined as follows. Events are selected online using a dimuon trigger with muon \pt thresholds
of 17 and 8\GeV. They must contain two oppositely charged muons with $\pt>20\GeV$ and $\abs{\eta}<2.4$,
both passing loose identification and isolation requirements.
The invariant mass of the dimuon system is required to be between 81 and 101\GeV.
In addition, the events are required to contain exactly one $\tauh$ candidate
passing the same selection requirements as in
the signal region, with $\pt^{\Pgt}>20\GeV$ and $\abs{\eta^{\Pgt}} < 2.1$.
To remove the overlap between muon and $\tauh$ candidates, the separation between them must fulfill
$\DR( \Pgm , \tauh)>0.1$, where $\DR$ is defined as $\DR =  \sqrt{\smash[b]{(\Delta\eta)^2+(\Delta\phi)^2}}$.
Data and simulation are compared using distributions of the
dimuon mass, \ptmiss, \pt/\ptmiss, \mT,  $\eta^{\Pgt}$ and $\pt^{\Pgt}$.
Figure~\ref{fig:CRplot} shows the $\pt^{\Pgt}$ distribution in the control region.
Data and simulation agree within 50\% in all bins except in one bin in the tail of the $\pt^{\Pgt}$ distribution, giving confidence that
the misidentified $\tauh$ background source---about 22\% of the total background---is correctly modeled in the
simulation.

\section{Systematic uncertainties}
\label{sec:uncertainties}

The uncertainty in the modeling of the \mT distribution can be split into three categories: uncertainties affecting shape and normalization, uncertainties affecting
only normalization and
an uncertainty due to limited numbers of events in simulated samples.

The dominant uncertainty of the first category comes
from $\tauh$ reconstruction and identification,
affecting background and a potential signal in the same way.
The uncertainty associated with the $\tauh$ identification is 5\%~\cite{Khachatryan:2015dfa}.
An additional systematic uncertainty, which dominates for high-$\pt$ $\tauh$ candidates, is related to the degree of confidence that the
MC simulation correctly models the identification
efficiency. This additional uncertainty increases linearly
with $\pt^{\Pgt}$ and amounts to $+5\%$/$-35\%$ at $\pt^{\Pgt} = 1\TeV$.
The uncertainty is asymmetric because studies
indicate that the $\Pgt$ identification efficiency is smaller in data than in simulation, and the difference increases as the \pt of the $\Pgt$ increases.
The uncertainty in the $\tauh$ energy scale amounts to 3\%~\cite{Khachatryan:2015dfa}.
The main sources of \ptmiss uncertainty from jets are the jet energy scale and jet energy resolution~\cite{CMS:2016ljj}.
For the energy measurements of other objects the following uncertainties are applied:
3\%~\cite{Khachatryan:2015dfa} for $\tauh$, 0.6\% in EB and 1.5\% in EE, respectively, for electrons and photons~\cite{Khachatryan:2015iwa};
and 0.2\% for muons~\cite{Sirunyan:2018fpa}.
The contribution to the uncertainty in \ptmiss associated with unclustered energy is estimated by varying this energy by $\pm$10\%.
For the $\Pgt$ plus \ptmiss trigger, a scale factor of 0.9 is applied. The scale factor has an uncertainty of 10\%.
The uncertainty associated with the choice of the PDF in the simulation is evaluated according to the PDF4LHC
prescription~\cite{Botje:2011sn,Alekhin:2011sk,Butterworth:2015oua}. The values increase with \mT,
ranging from an uncertainty of 1 to 10\% at $\mT=0.5$ to 4.0\TeV.
For the $K$ factor of the \PW\ boson background,
the difference between additive and multiplicative combination, which is around 5\%,
is taken to be the systematic uncertainty.
The simulated events are weighted so that the pileup distribution matches the measured one,
using a value for the total inelastic cross section of 69.2\unit{mb}, which has an uncertainty of $\pm$4.6 \%~\cite{Sirunyan:2018nqx}.

Uncertainties of the second category influence only the normalization of the \mT distribution.
Kinematic distributions in the $\PZ (\to \Pgm \Pgm)+\text{jets}$ control region
demonstrate that data and simulation agree within 50\% for misidentified $\tauh$ background, which is composed of $\PZ(\to \Pgn\Pgn)+\text{jets}$
and QCD multijet
events. This guides the assignment of a 50\% systematic uncertainty in the normalization of these backgrounds.
The uncertainty in the electron identification efficiency (veto) is 2\% and
the uncertainty in the integrated luminosity measurement is 2.5\%~\cite{LUM-17-001}.

Uncertainties in the third category arise from
limited sizes of event samples in the simulation
of background processes. In contrast to all other uncertainties, they are not correlated
between the bins of the invariant mass distribution.

In the high-mass region, where both the expected and the observed numbers of events are consistent with zero, the effect of the systematic uncertainty on the exclusion
limits is negligible.

The relevant systematic uncertainties taken into account in the estimation of potential signals
include those associated with $\tauh$ identification and energy scale, \ptmiss, trigger, pile-up simulation, and integrated luminosity.
The uncertainty
in the signal $K$ factor arises from the choices of PDF and \alpS. The combined uncertainty
is evaluated using the PDF4LHC prescription, where in the computation of each PDF set,
the strong coupling constant is varied. Uncertainties from different PDF sets and $\alpha_S$ variation
are added in quadrature.

\section{Results}
\label{sec:results}
The transverse mass distribution in Fig.~\ref{fig:MT} shows no significant deviations from the expected SM background.
Signal events
are expected to be particularly prominent at the upper end of the \mT
distribution, where the expected SM background is low.
The expected and measured yields are summarized in Table~\ref{Tab:numEvt}
together with the detailed systematic
uncertainties described in Section~\ref{sec:uncertainties}.

\begin{table*}[htp]
\topcaption{
Expected yields for the signal and background events compared to the measured event yields in data, for three regions of \mT.
Also shown are the total systematic uncertainties in the estimate of the event numbers.
}
\label{Tab:numEvt}
\centering
  \begin{tabular}{c  r@{}c@{}l  r@{}c@{}l  r@{}c@{}l }
    Range of \mT                     & \multicolumn{3}{l}{$\mT<0.5\TeV$} &  \multicolumn{3}{l}{$0.5<\mT<1\TeV$} &  \multicolumn{3}{l}{$\mT>1\TeV$}   \\  \hline
    $\PW+\text{jets}$                &  786 &  $\pm$ & 110     &  355   & $\pm$ & 68                 &  21.8   & $\pm$  & 6.2 \\
    $\PZ \to \Pgn\Pgn+\text{jets}$   &  238 &  $\pm$ & 120     &   68   & $\pm$ & 35                 &  0.9    & $\pm$  & 0.5  \\
    Multijet                         &  68  &  $\pm$ & 35      &  18    & $\pm$ & 10                 &         & $<$0.1 &  \\
    \ttbar                           &  68  &  $\pm$ & 15      &  14.5  & $\pm$ & 4.5                &         & $<$0.1 &  \\
    $\PZ \to \ell\ell + \text{jets}$ &  35.8 & $\pm$ & 8.7     &  10.4  & $\pm$ & 5.1                &         & $<$0.1 &  \\
    Diboson (\PW\PW, \PW\PZ, \PZ\PZ) &  24.9 & $\pm$ & 6.4     &  9.6   & $\pm$ & 3.5                &  0.7    & $\pm$  &  0.1  \\
    Single top quark                 &  21.5 & $\pm$ & 6.5     &  7.0   & $\pm$ & 2.9                &         & $<$0.1 & \\
    Total background                 &  1243 & $\pm$ & 160     &  485   & $\pm$ & 77                 &  23.4   & $\pm$  & 6.2  \\
[\cmsTabSkip]
    SSM \PWpr $M=600\GeV$          &  28229 & $\pm$ &  4388  &  14012 & $\pm$ & 2798               &  45.6   & $\pm$  & 7.7  \\
    SSM \PWpr $M=1\TeV$            &  3767 & $\pm$ & 590     &  10079 & $\pm$ & 1581               &  355    & $\pm$  & 98  \\
    SSM \PWpr $M=4\TeV$            &  0.7  & $\pm$ & 0.1     &  3.0   & $\pm$ & 1.8                &  11.4   & $\pm$  & 3.9  \\
[\cmsTabSkip]
    Data                             &       & 1203  &         &        & 452   &                    &         & 15     & \\  \hline
  \end{tabular}
\end{table*}

\subsection{Statistical analysis}

Upper limits on the product of the production cross section and branching fraction, $\sigma(\Pp\Pp \to \PWpr) \mathcal{B}(\PWpr \to \Pgt \Pgn)$,
are determined using a Bayesian method~\cite{stat,PDG2016} with a uniform positive 
prior probability density for the signal cross section (known to have excellent frequentist properties when used as a 
technical device for generating frequentist upper limits).
All limits presented here are at 95\% confidence level (\CL).
The nuisance parameters associated with the systematic uncertainties are modeled through log-normal distributions for uncertainties
in the normalization. Uncertainties in the shape of the distributions are modeled through
``template morphing'' techniques~\cite{Baak:2014fta}.
The limits are obtained from the entire \mT spectrum for $\mT>320\GeV$, as displayed in Fig.~\ref{fig:MT}.
This procedure is performed for different values of parameters of each signal, to obtain limits in terms on these parameters, such as the \PWpr boson mass.

To determine a model-independent upper limit on the product of the cross section and branching fraction, all events above a threshold $\mT^{\text{min}}$
are summed. From the number of background events, signal events, and observed data events, the cross section limit can be calculated.
The resulting limit can be reinterpreted
in the framework of
other models with a $\tauh$ and \ptmiss in the final state.

\subsection{The sequential standard model \PWpr}

The parameter of interest is the product of the signal cross section and the
branching fraction, $\sigma \mathcal{B}(\PWpr \to \Pgt \Pgn)$. The branching fraction
includes all $\Pgt$ lepton decay modes, to allow a direct comparison with the \PWpr searches in the
electron and muon channels~\cite{Sirunyan:2018mpc}.
\begin{figure}[htp]
\centering
\includegraphics[width=0.49\textwidth]{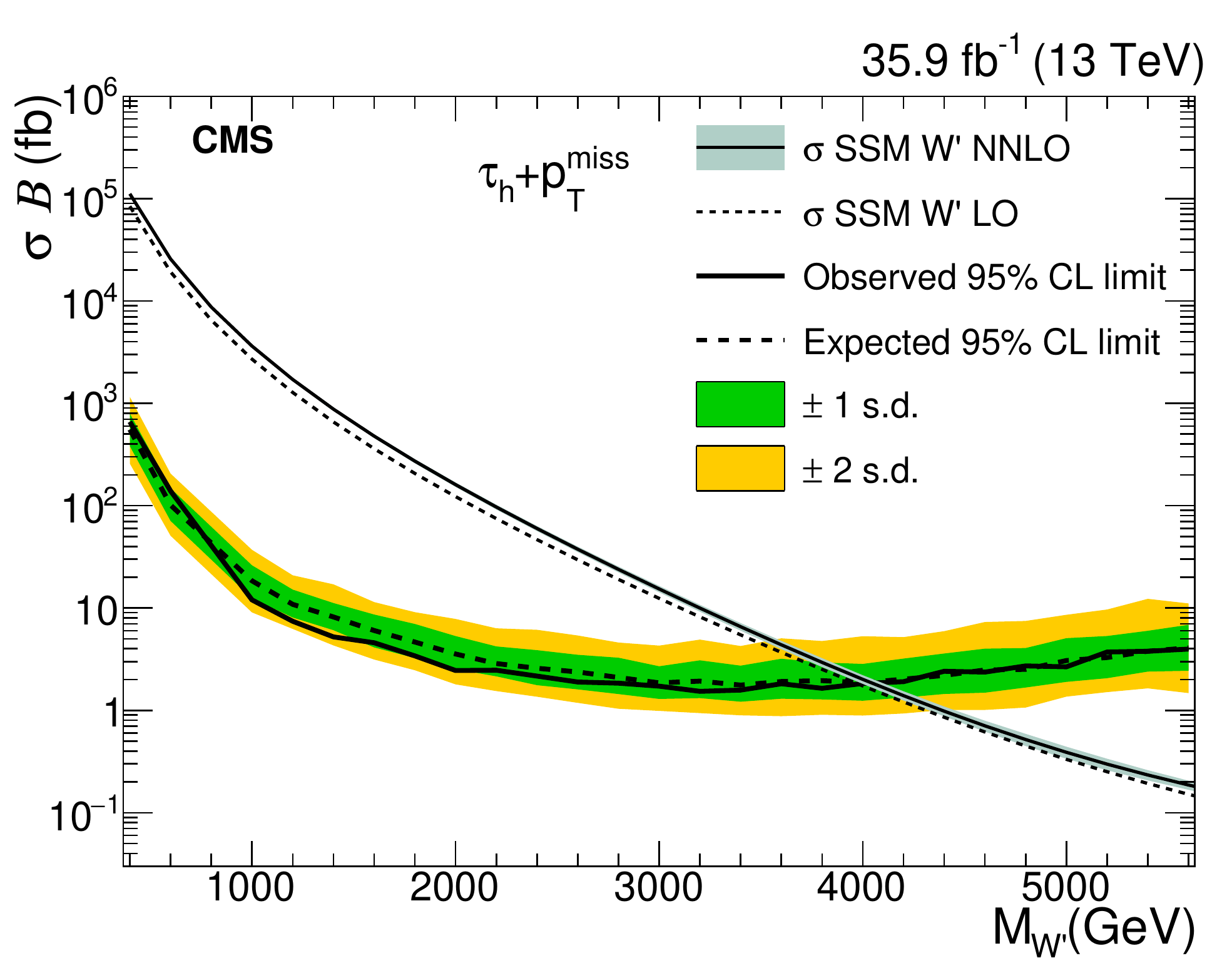}
\caption{Expected (black dashed line) and observed (black solid line) 95\% \CL upper limits on the cross section for the production of SSM \PWpr boson.
The shaded bands around the expected limit represent the one and two standard deviation (s.d.) uncertainty intervals.
The NNLO theoretical cross section with the corresponding PDF uncertainty band is also shown.
}
\label{fig:ssm_lim}
\end{figure}

The upper limit on $\sigma \mathcal{B}(\PWpr \to \Pgt \Pgn)$ as a function of the SSM \PWpr boson mass is shown in Fig.~\ref{fig:ssm_lim}.
The observed limit is consistent with the expected limit. The SSM \PWpr boson is excluded for masses
$0.4 < M_{\PWpr} < 4.0\TeV$ at 95\% \CL where the lower limit is mainly determined by the trigger threshold and the upper one
by the available data.
This result in the $\Pgt$ channel may be compared with the lower mass limit of 5.2\TeV for an SSM \PWpr boson, obtained from the
combination of electron and muon channels~\cite{Aaboud:2017efa,Sirunyan:2018mpc}.

\subsection{Limits on the coupling strength}

The upper limits on the cross section depend not only on the mass of a potential excess, but also
on the width.
Because of the relation between the coupling of a particle and its width,
a limit can also
be set on the coupling strength. In order to compute the limit for couplings $g_{\PWpr}/g_{\PW} \neq 1$, reweighted samples are
used that take into account the appropriate signal width and the differences in reconstructed
\mT shapes. For $g_{\PWpr}/g_{\PW} = 1$ the theoretical LO cross sections apply.
For a given mass, the cross section limit as a function of the coupling strength $g_{\PWpr}/g_{\PW}$ is determined.

\begin{figure}[htb]
\centering
\includegraphics[width=0.49\textwidth]{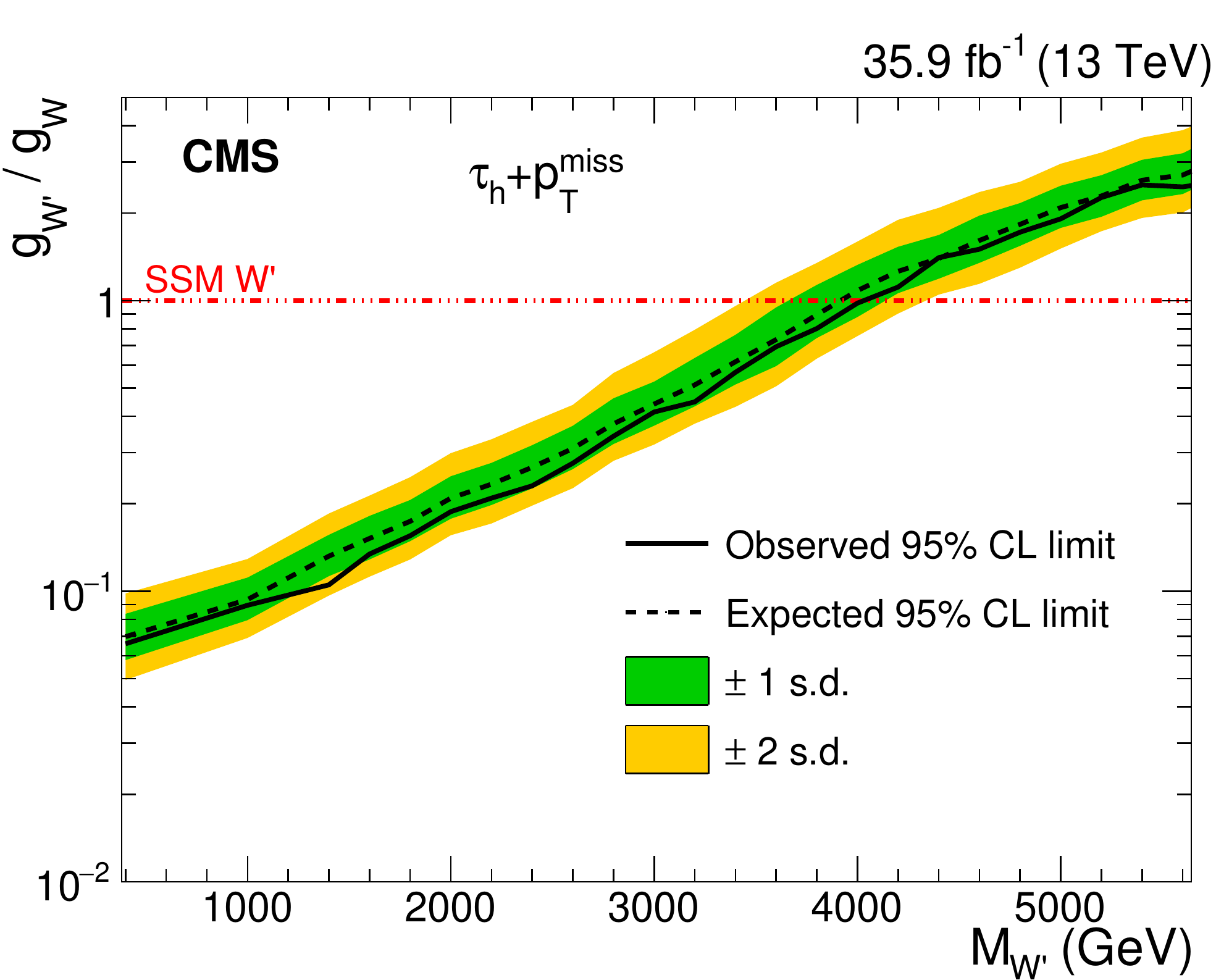}
\caption{
Expected (black dashed line) and observed (black solid line) 95\% \CL upper limits on the ratio of couplings as a function of the \PWpr boson mass.
The values above the observed limit contour are excluded.
The shaded bands around the expected limit represent the one and two standard deviation (s.d.) uncertainty intervals.
}
\label{fig:gc_lim}
\end{figure}
For each simulated \PWpr boson mass, the excluded cross section is determined from the intersection
of the theoretical cross section curve with the observed cross section limit.
The resulting intersection points provide the input for the exclusion limit in a two-dimensional plane
made of $g_{\PWpr}/g_{\PW}$ and $M_{\PWpr}$, as
depicted in Fig.~\ref{fig:gc_lim}. The phase space above the observed limit contour is excluded.
For low masses, $g_{\PWpr}/g_{\PW}$ values down to $7 \times 10^{-2}$ are excluded.

\subsection{The nonuniversal gauge interaction model limits}

\begin{figure*}[htp]
\centering
\includegraphics[width=0.49\textwidth]{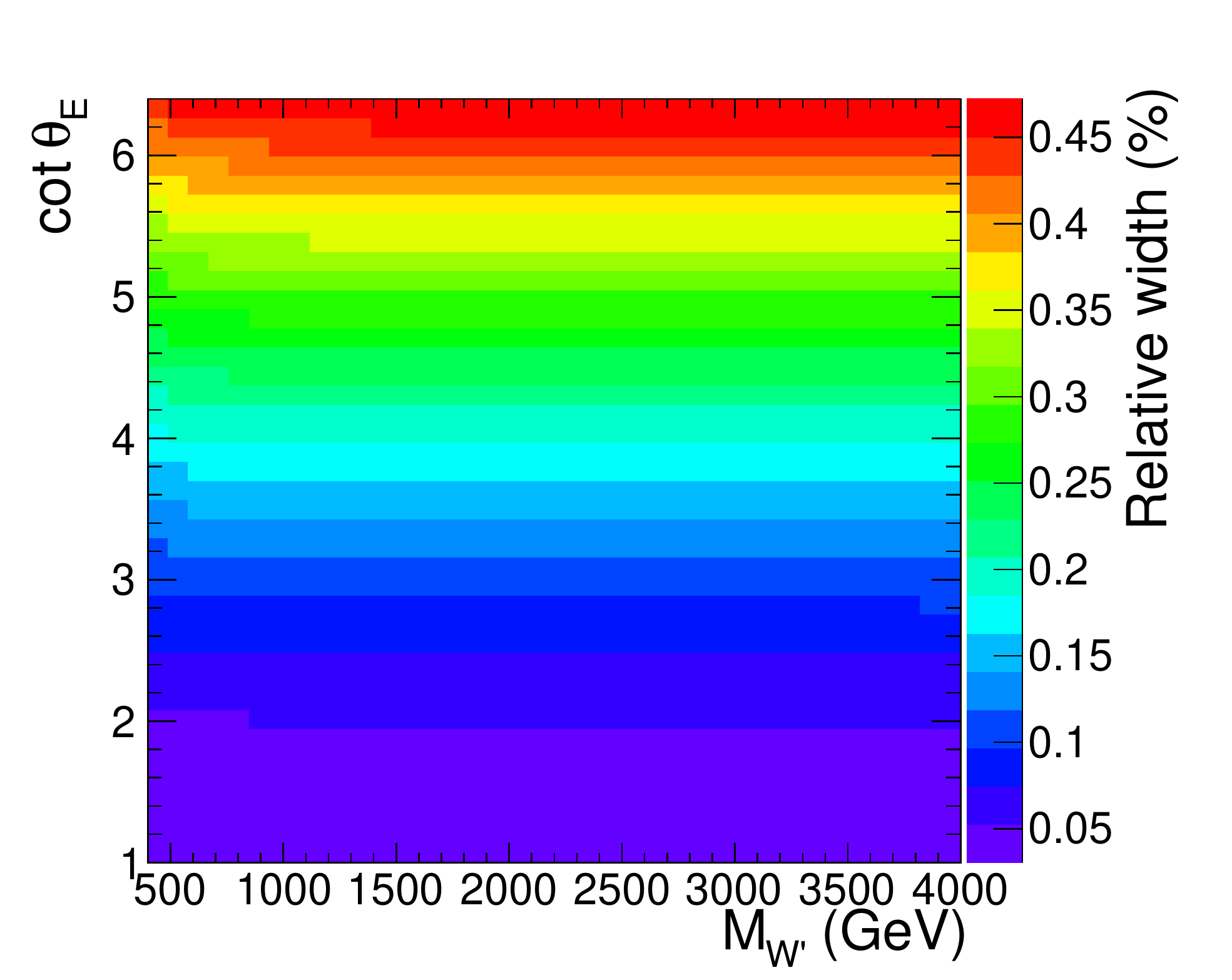}
\hskip 2 pt
\includegraphics[width=0.49\textwidth]{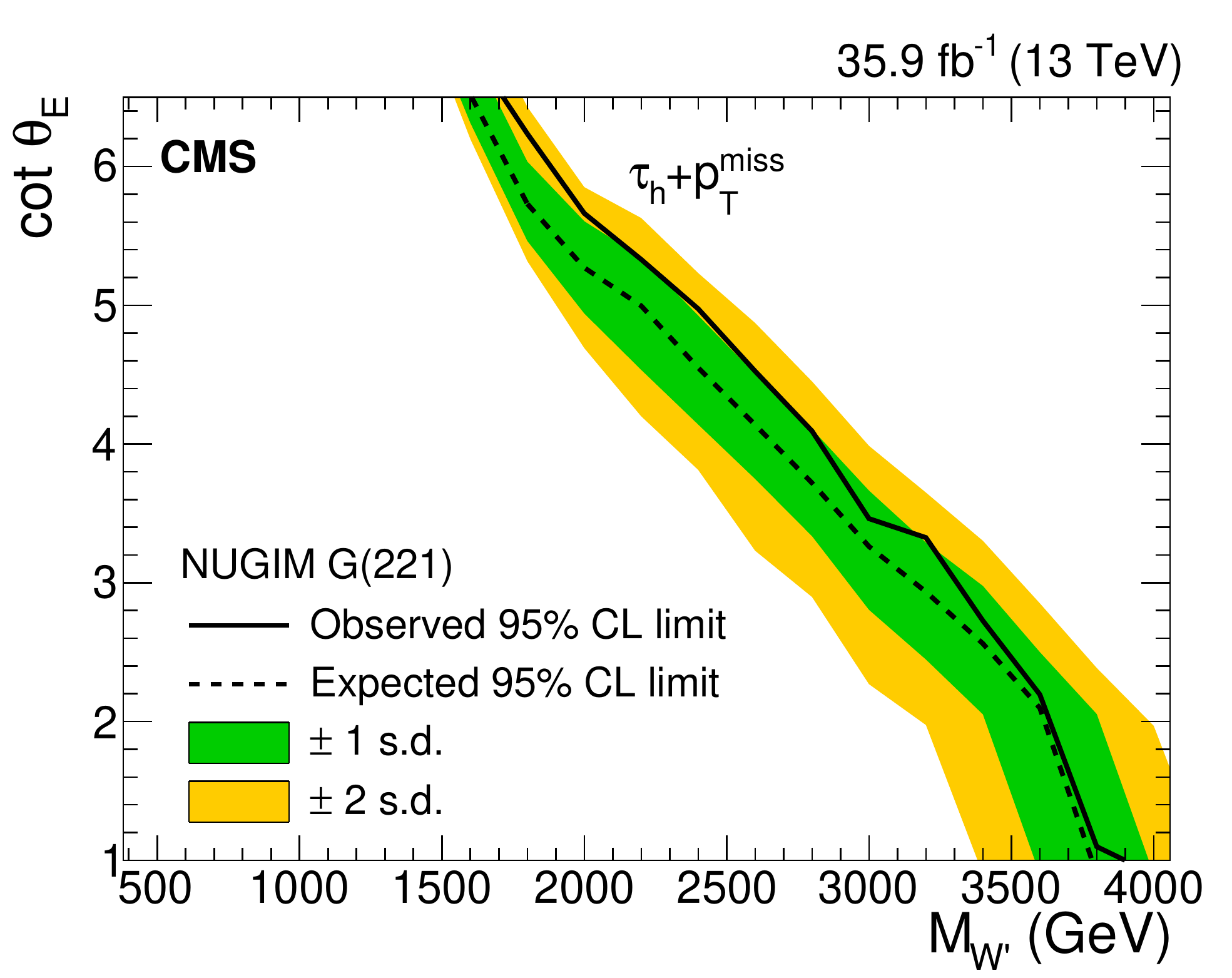}
\caption{
Left: The width of the $\PWpr$ boson as a function of $M_{\PWpr}$ and mixing angle $\cot\theta_{\mathrm{E}}$ in the NUGIM G(221) framework.
Right: Expected (black dashed line) and observed (black solid line) 95\% \CL
upper limits on the mixing angle $\cot \theta_{\mathrm{E}}$ as a function of the \PWpr boson mass.
The region left of the solid line is excluded.
The shaded bands represent the one and two standard deviation (s.d.) uncertainty bands.
}
\label{fig:NUGIM_lim}
\end{figure*}

In the NUGIM G(221) framework, the ratio of the couplings $g_{\PWpr}/g_{\PW}$ is related to the parameter $\cot\theta_{\mathrm{E}}$
through Eq.~\ref{eq1}.
Thus $\cot\theta_\mathrm{E}$ can be extracted for each value of $g_{\PWpr}/g_{\PW}$.
Based on the limits on coupling strengths presented in Fig.~\ref{fig:gc_lim}, the two-dimensional limit on $\cot\theta_\mathrm{E}$ is shown
as a function of the \PWpr boson mass.
Fig.~\ref{fig:NUGIM_lim} (left) shows the width of the \PWpr boson as a function of $\cot\theta_{\mathrm{E}}$ and $M_{\PWpr}$.
For $\cot\theta_{\mathrm{E}} > 6.5$, the width becomes so large that the model is no longer valid.
The limit, shown in Fig.~\ref{fig:NUGIM_lim} (right),
focuses on the parameter space $\cot\theta_{\mathrm{E}} \ge 1$
where the $\tauh$ channel sets the most stringent bounds, as illustrated in Fig.~\ref{fig:branching}.
For lower values of $\cot\theta_{\mathrm{E}}$,
other channels are more sensitive.
Depending on the value of
$\cot\theta_{\mathrm{E}}$, the mass of the \PWpr boson can be excluded at 95\% \CL
up to 3.9\TeV in the NUGIM G(221) framework.

\subsection{The model-independent cross section limit}

The shape analysis assumes a certain signal shape in \mT.
However, alternative new physics processes yielding a $\tauh+\ptmiss$ final state could cause an excess of a different shape.
A model-independent cross section limit is determined using a single bin ranging from a lower
threshold on \mT to infinity. No assumptions on the shape of the signal \mT distribution have
to be made other than that of a
flat product of acceptance times efficiency, $A \epsilon$, as a function of \PWpr mass.
In order to determine the limit for a specific model from the model-independent
limit shown here, only the model-dependent part of the efficiency needs to be applied. The
experimental efficiencies for the signal are already taken into account, including the effect of
the kinematic selection of events containing $\tauh$ and \ptmiss (the selections on $\pt/\ptmiss$ and $\Delta\phi$), the
geometrical acceptance (selection on $\eta$),
and the trigger threshold.

\begin{figure}[htb]
\centering
 \includegraphics[width=0.49\textwidth]{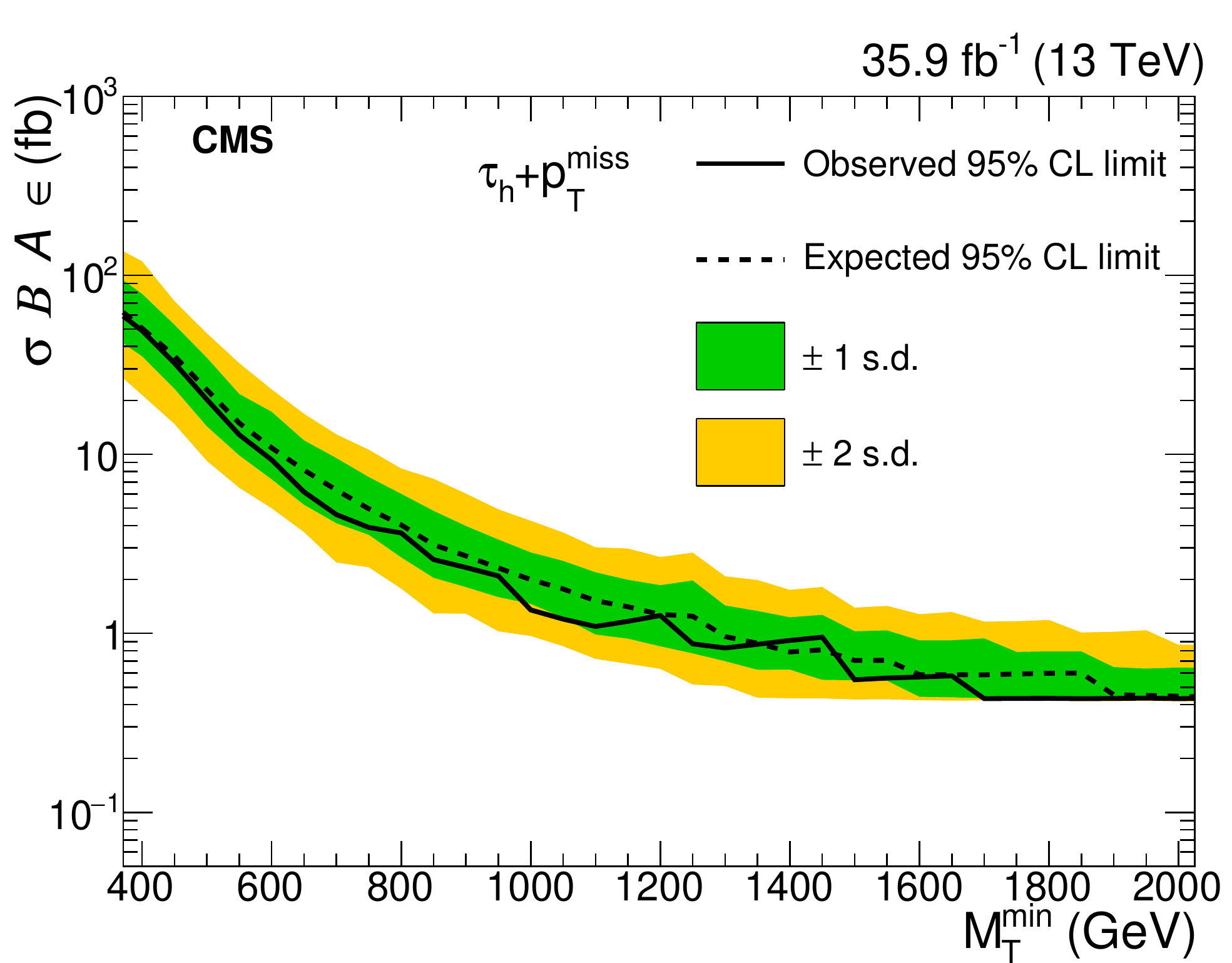}
 \caption{Expected (black dashed line) and observed (black solid line) 95\% \CL model-independent upper limits
on the product of cross section, branching fraction, and acceptance for a resonance decaying into the $\Pgt \Pgn$ channel.
The shaded bands represent the one and two standard deviation (s.d.) uncertainty bands.
}
\label{fig:mi_lim}
\end{figure}

A factor $f_{\mT}$ that reflects the effect of the threshold $\mT^{\text{min}}$ on the signal is determined by counting
the events with $\mT>\mT^{\text{min}}$ and dividing the result by the number of generated events. The reconstruction efficiency is nearly constant
over the entire \mT range probed here, therefore $f_{\mT}$ can be evaluated at generator level. A limit on the product of
the cross section and branching fraction $(\sigma \mathcal{B} A \epsilon)_{\text{excl}}$ can be obtained by dividing the excluded
cross section of the model-independent limit $(\sigma \mathcal{B} A \epsilon)_{\text{MI}}$ given in Fig.~\ref{fig:mi_lim} by the calculated fraction
$f_{\mT}(\mT^{\text{min}})$:\\
 \begin{equation}
(\sigma \mathcal{B} A \epsilon)_{\text{excl}} = \frac{(\sigma \mathcal{B} A \epsilon)_{\text{MI}}(\mT^{\text{min}})}{f_{\mT}(\mT^{\text{min}})}
 \end{equation}
Here, $\mathcal{B}$ is the branching fraction of the new particle decaying to $\Pgt +\Pgn$.
Models with a theoretical cross section $(\sigma \mathcal{B})_{\text{theo}}$ larger than $(\sigma \mathcal{B})_{\text{excl}}$ can be excluded.
The procedure described here can be applied to all models involving the two-body decay of a massive state,
which exhibit back-to-back kinematics similar to those of a generic \PWpr. If the kinematic properties are different,
the fraction of events $f_{\mT}(\mT^{\text{min}})$ must be determined for the particular model considered.

The resulting cross section limit as a function of $\mT^\text{min}$ is shown in Fig.~\ref{fig:mi_lim}.
The highest \mT event in data was found at 1.65\TeV, after which the limit becomes flat.
The results depend strongly on the threshold $\mT^{\text{min}}$.
Values of the product $\sigma \mathcal{B} A \epsilon$
between 50 fb ($\mT^{\text{min}}>400\GeV$) and 0.4 fb ($\mT^{\text{min}}>2\TeV$) are
excluded for the $\mT^{\text{min}}$ thresholds given in brackets.

\section{Summary}

A search for new physics in final states with a hadronically decaying $\Pgt$ lepton and missing transverse momentum has been performed
by the CMS experiment, using  proton-proton collision data at the center-of-mass energy $\sqrt{s}=13\TeV$
with an integrated luminosity of 35.9\fbinv. No significant excess
compared to the standard model expectation
is observed in the transverse mass of the $\Pgt$ and missing transverse momentum.
A sequential standard model \PWpr boson is excluded in the mass range $0.4 < M_{\PWpr} < 4.0\TeV$ at 95\% confidence level.
Couplings that are weaker than assumed in the sequential standard model can be excluded down to values of $7\times10^{-2}$ for $M_{\PWpr}=1\TeV$.
Within the nonuniversal gauge interaction $SU(2) \times SU(2) \times U(1)$ model,
the lower limit on the \PWpr boson mass depends on the
coupling constant
and varies
from 1.7 to 3.9\TeV at 95\% confidence level.
For $\cot\theta_{\mathrm{E}}>1$, these results obtained in the $\Pgt$ channel provide the most stringent constraints on this model to date.
In addition,
a model-independent limit
is provided allowing the results to be interpreted in other models giving the same final state with similar kinematic distributions.

\begin{acknowledgments}

We congratulate our colleagues in the CERN accelerator departments for the excellent performance of the LHC and thank the technical and administrative staffs at CERN and at other CMS institutes for their
contributions to the success of the CMS effort. In addition, we gratefully acknowledge the computing centers and personnel of the Worldwide LHC Computing Grid for delivering so effectively the computing
infrastructure essential to our analyses. Finally, we acknowledge the enduring support for the construction and operation of the LHC and the CMS detector provided by the following funding agencies: BMWFW and
FWF (Austria); FNRS and FWO (Belgium); CNPq, CAPES, FAPERJ, FAPERGS, and FAPESP (Brazil); MES (Bulgaria); CERN; CAS, MoST, and NSFC (China); COLCIENCIAS (Colombia); MSES and CSF (Croatia); RPF (Cyprus);
SENESCYT (Ecuador); MoER, ERC IUT, and ERDF (Estonia); Academy of Finland, MEC, and HIP (Finland); CEA and CNRS/IN2P3 (France); BMBF, DFG, and HGF (Germany); GSRT (Greece); NKFIA (Hungary); DAE and DST
(India); IPM (Iran); SFI (Ireland); INFN (Italy); MSIP and NRF (Republic of Korea); MES (Latvia); LAS (Lithuania); MOE and UM (Malaysia); BUAP, CINVESTAV, CONACYT, LNS, SEP, and UASLP-FAI (Mexico); MOS
(Montenegro); MBIE (New Zealand); PAEC (Pakistan); MSHE and NSC (Poland); FCT (Portugal); JINR (Dubna); MON, RosAtom, RAS, RFBR, and NRC KI (Russia); MESTD (Serbia); SEIDI, CPAN, PCTI, and FEDER (Spain);
MOSTR (Sri Lanka); Swiss Funding Agencies (Switzerland); MST (Taipei); ThEPCenter, IPST, STAR, and NSTDA (Thailand); TUBITAK and TAEK (Turkey); NASU and SFFR (Ukraine); STFC (United Kingdom); DOE and NSF
(USA).

\hyphenation{Rachada-pisek} Individuals have received support from the Marie-Curie programme and the European Research Council and Horizon 2020 Grant, contract No. 675440 (European Union); the Leventis
Foundation; the A. P. Sloan Foundation; the Alexander von Humboldt Foundation; the Belgian Federal Science Policy Office; the Fonds pour la Formation \`a la Recherche dans l'Industrie et dans l'Agriculture
(FRIA-Belgium); the Agentschap voor Innovatie door Wetenschap en Technologie (IWT-Belgium); the F.R.S.-FNRS and FWO (Belgium) under the ``Excellence of Science - EOS" - be.h project n. 30820817; the Ministry
of Education, Youth and Sports (MEYS) of the Czech Republic; the Lend\"ulet (``Momentum") Programme and the J\'anos Bolyai Research Scholarship of the Hungarian Academy of Sciences, the New National
Excellence Program \'UNKP, the NKFIA research grants 123842, 123959, 124845, 124850 and 125105 (Hungary); the Council of Science and Industrial Research, India; the HOMING PLUS programme of the Foundation for
Polish Science, cofinanced from European Union, Regional Development Fund, the Mobility Plus programme of the Ministry of Science and Higher Education, the National Science Center (Poland), contracts Harmonia
2014/14/M/ST2/00428, Opus 2014/13/B/ST2/02543, 2014/15/B/ST2/03998, and 2015/19/B/ST2/02861, Sonata-bis 2012/07/E/ST2/01406; the National Priorities Research Program by Qatar National Research Fund; the
Programa Estatal de Fomento de la Investigaci{\'o}n Cient{\'i}fica y T{\'e}cnica de Excelencia Mar\'{\i}a de Maeztu, grant MDM-2015-0509 and the Programa Severo Ochoa del Principado de Asturias; the Thalis
and Aristeia programmes cofinanced by EU-ESF and the Greek NSRF; the Rachadapisek Sompot Fund for Postdoctoral Fellowship, Chulalongkorn University and the Chulalongkorn Academic into Its 2nd Century Project
Advancement Project (Thailand); the Welch Foundation, contract C-1845; and the Weston Havens Foundation (USA).

\end{acknowledgments}
\clearpage
\bibliography{auto_generated}

\cleardoublepage \appendix\section{The CMS Collaboration \label{app:collab}}\begin{sloppypar}\hyphenpenalty=5000\widowpenalty=500\clubpenalty=5000\vskip\cmsinstskip
\textbf{Yerevan Physics Institute, Yerevan, Armenia}\\*[0pt]
A.M.~Sirunyan, A.~Tumasyan
\vskip\cmsinstskip
\textbf{Institut f\"{u}r Hochenergiephysik, Wien, Austria}\\*[0pt]
W.~Adam, F.~Ambrogi, E.~Asilar, T.~Bergauer, J.~Brandstetter, M.~Dragicevic, J.~Er\"{o}, A.~Escalante~Del~Valle, M.~Flechl, R.~Fr\"{u}hwirth\cmsAuthorMark{1}, V.M.~Ghete, J.~Hrubec, M.~Jeitler\cmsAuthorMark{1}, N.~Krammer, I.~Kr\"{a}tschmer, D.~Liko, T.~Madlener, I.~Mikulec, N.~Rad, H.~Rohringer, J.~Schieck\cmsAuthorMark{1}, R.~Sch\"{o}fbeck, M.~Spanring, D.~Spitzbart, A.~Taurok, W.~Waltenberger, J.~Wittmann, C.-E.~Wulz\cmsAuthorMark{1}, M.~Zarucki
% ...
\vskip\cmsinstskip
% ...
1:  Also at Vienna University of Technology, Vienna, Austria\\
% ...
\end{sloppypar}
\end{document}